\documentclass[12pt,twoside]{article}
\usepackage{amsmath,amsfonts,amssymb}
\usepackage{graphicx}
\usepackage{eucal}
\usepackage{mathrsfs}
\usepackage{theorem}
\usepackage{pifont}
\newcommand{\Real}{\mathbb{R}}

\newcommand{\todo}[1]{{\sffamily To do:}}

\newtheorem{theorem}{Theorem}

\newtheorem {lemma}{Lemma}
\newenvironment{proof}{{\flushleft \emph{Proof}:}}{\ding{110}}

\date{}
\title{Existence of periodic solutions for enzyme-catalysed reactions with periodic substrate input}
\author{Guy Katriel\footnotemark[1]}

\begin{document}
\maketitle
\begin{abstract}
Considering a basic enzyme-catalysed reaction, in which the rate of
input of the substrate varies periodically in time, we give a
necessary and sufficient condition for the existence of a periodic
solution of the reaction equations. The proof employs the
Leray-Schauder degree, applied to an appropriately constructed
homotopy.
\end{abstract}

\renewcommand{\thefootnote}{\fnsymbol{footnote}}
\footnotetext[1]{Institute of Mathematics, The Hebrew University,
Jerusalem 91904, Israel.

Partially supported by the Edmund Landau Center for Research in
Mathematical Analysis and Related Areas, sponsored by the Minerva
Foundation (Germany).}

\section{Introduction}
The basic scheme for a reaction catalysed by an enzyme is
$$\begin{array}{c}
                                    I(t)\\
                                      \rightarrow\\
                                      \;
                                    \end{array}
                                    S+E\begin{array}{c}
                                      k_1\\
                                      \rightleftarrows\\
                                      k_{-1}
                                    \end{array}
                                    C
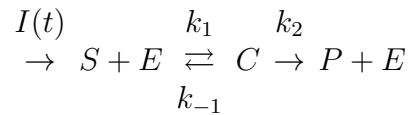
                                    \begin{array}{c}
                                    k_2\\
                                      \rightarrow\\
                                      \;
                                    \end{array}
                                    P+E
$$
in which the substrate $S$ and the enzyme $E$ form a complex $C$
through a reversible reaction, and the complex $C$ can dissociate
into the enzyme and the product $P$. $I(t)$ is the rate of input of
the substrate into the system, satisfying
\begin{equation}\label{posi}
I(t)\geq 0,\;\;\; t\in\Real.
\end{equation}

The dynamics of this system are described by the rate equations for
the concentrations of the species:
\begin{equation}\label{e1}E'(t)=-k_1 E(t)S(t)+(k_{-1}+k_2)C(t)
\end{equation}
\begin{equation}\label{e2}S'(t)=I(t)-k_1 E(t)S(t)+k_{-1}C(t)
\end{equation}
\begin{equation}\label{e3}C'(t)=k_1 E(t)S(t)-(k_{-1}+k_2)C(t)
\end{equation}
\begin{equation}\label{e4}P'(t)=k_2 C(t)
\end{equation}

Stoleriu, Davidson and Liu \cite{stoleriu} recently investigated the
case in which the rate of input of the substrate, $I(t)$, fluctuates
in time in a periodic manner. As they noted, such a situation is
common in biological systems, both due to intrinsic oscillations in
preceding steps of the reaction pathway, and to oscillations
external to the organism.

Since by adding (\ref{e1}) and (\ref{e3}) we have $[E(t)+C(t)]'=0$,
that is the total amount of the free and bound enzyme is constant in
time, we can set
$$E(t)+C(t)=K,$$
and since (\ref{e4}) decouples from the other equations, we can
rewrite the system in terms of $S$ and $C$ as
\begin{equation}\label{de1}
S'(t)=I(t)-k_1[K-C(t)]S(t)+k_{-1}C(t),
\end{equation}
\begin{equation}\label{de2}
C'(t)=k_1[K-C(t)]S(t)-(k_{-1}+k_2)C(t).
\end{equation}
Assuming that $I(t)$ satisfies (\ref{posi}) and is $T$-periodic
\begin{equation}\label{per}
I(t+T)=I(t),\;\;\; t\in\Real,
\end{equation}
we ask, as a first step in understanding the dynamics of the system,
whether there exists a $T$-periodic solution $S(t),C(t)$, of
(\ref{de1}), (\ref{de2}), with
\begin{equation}\label{pos1}
S(t)>0,  \;\;\;  t\in\Real,
\end{equation}
\begin{equation}\label{pos2}
0<C(t)<K,  \;\;\; t\in\Real.
\end{equation}

In the case of a constant rate of input $I(t)=\bar{I}$, there is a
unique stationary solution if and only if
\begin{equation}\label{cc}
\bar{I}<k_2 K,
\end{equation}
given by
\begin{equation}\label{stat}
\bar{C}=\frac{1}{k_2}\bar{I},\;\;\;\;\bar{S}=\frac{(k_2+k_{-1})\bar{I}}{k_1(k_2K-\bar{I})},\end{equation}
and a phase-plane analysis shows that all solutions tend to this
equilibrium. Note that the condition (\ref{cc}) says that the rate
of input $\bar{I}$ of the substrate is not too large - indeed if
this condition is violated then $S(t)$ will increase without bound,
since the substrate enters the system more rapidly than it can be
processed by the enzyme. The proof of our existence theorem for the
periodic case will involve a homotopy connecting it to the case of
constant input.

Returning to the general case of periodic $I(t)$, and adding
(\ref{de1}) and (\ref{de2}) we get
\begin{equation}\label{del}
C'(t)+k_2C(t)=I(t)-S'(t).
\end{equation}
Assuming that $C(t),S(t)$ is a $T$-periodic solution, and
integrating (\ref{del}) on $[0,T]$, taking into account the
periodicity, we get
\begin{equation}\label{av0}
k_2 \int_0^T  C(t)dt = \int_0^T  I(t)dt.
\end{equation}
From (\ref{pos2}) and (\ref{av0}) it follows that
\begin{equation}\label{ned}
\frac{1}{T}\int_0^T  I(t)dt<k_2 K.
\end{equation}
Thus (\ref{ned}) is a necessary condition for the existence of a
$T$-periodic solution.

In \cite{stoleriu}, (where the case $I(t)=\bar{I}(1+\epsilon
\sin(\omega t))$ with $0\leq \epsilon\leq 1$ is considered) it is
proposed that the existence of a periodic solution can be proven by
constructing an invariant rectangle for the flow corresponding to
the system (\ref{de1}),(\ref{de2}), of the form
\begin{equation}\label{defd} D=\{ (S,C)\;|\; 0\leq S\leq
\hat{S},\;\; 0\leq C\leq K-\delta \},\end{equation} (with $\hat{S}$
and $\delta$ chosen appropriately), so that Brouwer's fixed-point
theorem, applied to the time $T$ Poincar\'e map of
(\ref{de1}),(\ref{de2}), implies the existence of a fixed point of
this map, which corresponds to the required $T$-periodic solution.
An examination of this method of proof shows that an invariant
rectangle of the form (\ref{defd}) exists {\it{if and only if}} the
input function $I(t)$ satisfies
\begin{equation}\label{st}
\max_{t\in [0,T]} I(t) <k_2 K.
\end{equation}
Note that this sufficient condition (\ref{st}) for the existence of
a $T$-periodic solution is more stringent than the necessary
condition (\ref{ned}). Here we bridge this gap by proving that in
fact (\ref{ned}) is sufficient for the existence of a positive
$T$-periodic solution. We employ the methods of nonlinear functional
analysis, reformulating the problem as a fixed-point problem for a
nonlinear operator in a space of $T$-periodic functions, and using
degree theory (see, {\it{e.g.}}, \cite{brown,zeidler}) to prove the
existence of a fixed point of this operator.

Our existence result raises the following question: is it true that
for any $I(t)$ satisfying (\ref{ned}), the periodic solution is
{\it{unique}} and {\it{globally stable}}?

\section{The existence theorem}

We set
$$\bar{I}=\frac{1}{T}\int_0^T
I(t)dt,\;\;\;\;I_0(t)=I(t)-\bar{I}.$$

\begin{theorem}\label{main}
Assume $k_1,k_2>0$, $k_{-1}\geq 0$, $K>0$ and $I(t)$ is a continuous
function satisfying (\ref{posi}) and (\ref{per}). Then there exists
a $T$-periodic solution $S(t),C(t)$ of (\ref{de1}),(\ref{de2})
satisfying (\ref{pos1}),(\ref{pos2}) if and only if
\begin{equation}
\label{ec}0<\bar{I}<k_2 K.
\end{equation}
\end{theorem}

The fact that (\ref{ec}) is a necessary condition for existence was
explained above, so we now assume that (\ref{ec}) holds, and need to
prove the existence of a $T$-periodic solution.

Note that (\ref{de1}),(\ref{de2}) is equivalent to
(\ref{de1}),(\ref{del}), and it is the latter which we will be
using.

We define
\begin{equation}\label{dcb}\bar{C}=\frac{1}{k_2}\bar{I},\end{equation}
$$C_0(t)=C(t)-\bar{C}.$$
The condition (\ref{ec}) is equivalent to
\begin{equation}
\label{ec1}0<\bar{C}< K,
\end{equation}
and from (\ref{av0}) we have that if $S,C$ is a $T$-periodic
solution of (\ref{de1}),(\ref{del}) then $C_0$ satisfies
\begin{equation}\label{c0} \int_0^T C_0(t)dt =0.
\end{equation}

We can thus rewrite (\ref{de1}),(\ref{del}) in terms of $S$ and
$C_0$:
\begin{equation}\label{de1A}
S'(t)+k_1[K-\bar{C}-C_0(t)]S(t)=I(t)+k_{-1}C_0(t)+k_{-1}\bar{C},
\end{equation}\begin{equation}\label{delA}
C_0'(t)+k_2C_0(t)=I_0(t)-S'(t).
\end{equation}

The following lemma, in which we solve (\ref{de1A}) for a
$T$-periodic solution $S(t)$ in terms of $C_0(t)$, is the place
where the key condition (\ref{ec}) is exploited.

\begin{lemma}\label{ex} Given a $T$-periodic continuous function $C_0(t)$
satisfying (\ref{c0}), the linear equation (\ref{de1A}) has a unique
$T$-periodic solution $S(t)$, and if, in addition,
\begin{equation}\label{nnn}
C(t)=\bar{C}+C_0(t)\geq 0,\;\;\; t\in\Real,
\end{equation}
then $S$ satisfies (\ref{pos1}).
\end{lemma}

\begin{proof}
Setting $C(t)=\bar{C}+C_0(t)$, the general solution of (\ref{de1A})
is
\begin{eqnarray}\label{dd}
S(t)&=&\exp \Big(-k_1 \int_0^t [K-C(s)]ds \Big)S(0)\\&+& \int_0^t
\exp \Big(k_1\int_t^s [K-C(r)]dr \Big)[I(s)+k_{-1}C(s)] ds.\nonumber
\end{eqnarray}
The condition for $S$ to be $T$-periodic is $S(0)=S(T)$, or
\begin{eqnarray}\label{tt}
\Big[1&-&\exp \Big(-k_1 \int_0^T [K-C(s)] ds \Big)
\Big]S(0)\\&=&e^{k_1 T[K-\bar{C}]}\int_0^T \exp \Big(k_1 \int_0^s
[K-C(r)]dr \Big)[I(s)+k_{-1}C(s)]ds.\nonumber
\end{eqnarray}
(\ref{ec1}),(\ref{dcb}) and (\ref{c0}) imply that
$$\int_0^T [K-C(t)] ds = T[K- \bar{C}] >0,$$
which implies that the coefficient of $S(0)$ on the left-hand side
of (\ref{tt}) is positive, hence  the existence of a unique
$T$-periodic solution $S$ of (\ref{delA}) is ensured, given by
\begin{eqnarray}\label{defs}
&&S(t)=\int_0^t \exp \Big(k_1\int_t^s [K-C(r)]dr
\Big)[I(s)+k_{-1}C(s)] ds \\&+&[e^{k_1 T(K-
\bar{C})}-1]^{-1}\int_0^T \exp \Big( k_1\int_t^s
[K-C(r)]dr\Big)[I(s)+k_{-1}C(s)]ds.\nonumber
\end{eqnarray}
If (\ref{nnn}) holds, then recalling (\ref{posi}) and (\ref{ec1}),
we see that the second term in (\ref{defs}) is strictly positive,
and since the first term is nonnegative, we get (\ref{pos1}).
\end{proof}

We define $X$ to be the space of continuous $T$-periodic functions
$C_0(t)$ satisfying (\ref{c0}),
 with the maximum norm.

Fixing  $I(t)$ satisfying (\ref{ec}), we use the result of lemma
\ref{ex} to define a mapping ${\mathcal{F}}(I;\cdot):X\rightarrow X$
($I$ is regarded as a parameter) as follows. Let $C_0\in X$ and let
$S$ be the $T$-periodic solution of (\ref{de1A}) (given explicitly
by (\ref{defs})). Then define
$${\mathcal{F}}(I;C_0)=S'.$$
Note that since $S$ is periodic the integral of $S'$ over $[0,T]$ is
$0$, so we indeed have ${\mathcal{F}}(I;C_0)\in X$. From the
explicit formula (\ref{defs}) it can be be shown by standard methods
that
\begin{lemma}\label{pf} The mapping ${\mathcal{F}}(I;.):X\rightarrow
X$ is Fr\'echet differentiable, and maps bounded sets to bounded
sets.
\end{lemma}

We can now reformulate the problem of finding periodic solutions of
(\ref{de1A}),(\ref{delA}) as: find $C_0\in X$  satisfying
\begin{equation}\label{rf}
C_0'(t)+k_2C_0(t)=I_0(t)-{\mathcal{F}}(I;C_0)(t).
\end{equation}

We define a linear mapping ${\mathcal{L}}:X\rightarrow X$: for any
$R\in X$, define ${\mathcal{L}}(R)=C$ to be the unique $T$-periodic
solution of the equation
\begin{equation}\label{ww}
C'(t)+k_2C(t)=R(t).
\end{equation}
By integrating (\ref{ww}) over $[0,T]$ we see that
$$\int_0^T C(t)dt=0,$$
so that ${\mathcal{L}}(R)\in X$. Moreover, since in fact
${\mathcal{L}}$ maps $X$ boundedly to the space of $C^1$-functions,
which is compactly embedded in $X$, we have that

\begin{lemma}\label{comp}
${\mathcal{L}}:X\rightarrow X$ is {\it{compact}}.
\end{lemma}

We can now rewrite (\ref{rf}) as
\begin{equation}\label{op}
C_0={\mathcal{L}}(I_0-{\mathcal{F}}(I;C_0)).
\end{equation}

We will prove the existence of a solution of (\ref{op}) by applying
degree theory. Fixing $I$, we define the homotopy
${\mathcal{H}}:[0,1]\times X\rightarrow X$ by
$${\mathcal{H}}(\lambda,C_0)={\mathcal{L}}(\lambda I_0-{\mathcal{F}}(\bar{I}+\lambda I_0; C_0)),$$
and we will consider the equation
\begin{equation}\label{ho}
C_0={\mathcal{H}}(\lambda,C_0)\;\;\;0\leq \lambda \leq 1.
\end{equation}
Note that for $\lambda=1$ (\ref{ho}) coincides with (\ref{op}),
while for $\lambda=0$ (\ref{ho}) corresponds to the system with
constant rate $\bar{I}$ of substrate input.

From lemmas \ref{pf} and \ref{comp} we have
\begin{lemma}${\mathcal{H}}: [0,1]\times X\rightarrow X$ is a continuous compact
mapping.
\end{lemma}
This ensures that we can apply the Leray-Schauder degree (see,
{\it{e.g.}}, \cite{brown,zeidler}) to ${\mathcal{H}}$.

We define the bounded open set $G\subset X$ by
$$G=\{C_0\in X\;|\; 0<C_0(t)+\bar{C}<K \;\;\; t\in \Real \}.$$

The following lemma summarizes the reduction of our problem to a
fixed-point problem, and provides essential a-priori bounds.
\begin{lemma}\label{nn}
If $C_0\in \bar{G}$ solves (\ref{ho}) for some $0\leq \lambda \leq
1$, and if $C=C_0+\bar{C}$ and $S$ is the $T$-periodic solution of
\begin{equation}\label{de1s}
S'(t)=\bar{I}+\lambda I_0(t)-k_1[K-C(t)]S(t)+k_{-1}C(t),
\end{equation}
then
\begin{equation}\label{de2s}
C'(t)=k_1[K-C(t)]S(t)-(k_{-1}+k_2)C(t).
\end{equation}
and $S,C$ satisfy (\ref{pos1}),(\ref{pos2}).
\end{lemma}

\begin{proof}
Assume $C_0$ satisfies (\ref{ho}). By the definition of
${\mathcal{F}}$ we have ${\mathcal{F}}(\bar{I}+\lambda I_0; C_0)=S'$
where $S$ is the periodic solution of (\ref{de1s}), and by the
definition of ${\mathcal{L}}$ we have
\begin{equation}\label{ee1}C_0'(t)+k_2C_0(t)=\lambda
I_0(t)-{\mathcal{F}}(\bar{I}+\lambda I_0; C_0)=\lambda I_0(t)-S'(t).
\end{equation}
Taking the difference of (\ref{de1s}) and (\ref{ee1}) gives
(\ref{de2s}).

By (\ref{posi}) we have
\begin{equation}\label{sp}
\bar{I}+\lambda I_0(t)=(1-\lambda)\bar{I}+\lambda I(t)\geq 0, \;\;\;
\lambda\in [0,1],\; t\in\Real,
\end{equation}
and the assumption that $C_0\in\bar{G}$ means that
\begin{equation}\label{kk}
0\leq C(t) \leq K,\;\;\; t\in\Real,
\end{equation}
so lemma \ref{ex} implies that $S$ satisfies (\ref{pos1}).

To show that $C(t)>0$ for all $t$, let $t_1$ be a point where $C$
achieves its global minimum, and assume by way of contradiction that
$C(t_1)\leq 0$, so that by (\ref{kk}) $C(t_1)=0$. Since $t_1$ is a
minimum point, we also have $C'(t_1)=0$. Substituting into
(\ref{de2s}) we get $k_1 K S(t_1)=0$, so $S(t_1)=0$, which
contradicts (\ref{pos1}), which has already been proved.

Similarly, to show that $C(t)<K$ for all $t$, let $t_2$ be a point
where $C$ achieves its global maximum, and assume by way of
contradiction that $C(t_2)\geq K$, so that by (\ref{kk}) $C(t_2)=K$.
Since $t_2$ is a maximum point, we also have $C'(t_2)=0$.
Substituting into (\ref{de2s}) gives $(k_{-1}+k_2)K=0$, a
contradiction.
\end{proof}

We note that lemma \ref{nn} implies that if $C_0\in {\bar{G}}$ then
$C_0$ satisfies (\ref{pos2}), so that $C_0\in G$. We thus get

\begin{lemma} $\lambda\in [0,1], \;C_0\in \partial G\;\;\;\Rightarrow\;\;\;C_0\neq {\mathcal{H}}(\lambda,C_0).$
\end{lemma}
By the homotopy invariance property of the degree, it follows that
\begin{lemma}\label{hi}$deg(id_X-{\mathcal{H}}(1,\cdot),G)=deg(id_X-{\mathcal{H}}(0,\cdot),G).$
\end{lemma}

We now show that
\begin{lemma}\label{dnz}
$deg(id_X-{\mathcal{H}}(0,\cdot),G)\neq 0$.
\end{lemma}

\begin{proof}
When $\lambda=0$, the only $T$-periodic solution of
(\ref{de1s}),(\ref{de2s}) is that given by (\ref{stat}), so the only
solution of (\ref{ho}) is $C_0=0$.

In order to prove the lemma it suffices, then, to prove that the
local degree of $id_X-{\mathcal{H}}(0,\cdot)$ at $C_0=0$ is nonzero,
and this will follow if we can show the Fr\'echet derivative
$id_X-D_{C_0}{\mathcal{H}}(0,0)$ is nonsingular.

We have
$$id_X-D_{C_0}{\mathcal{H}}(\bar{I},0)=id_X+{\mathcal{L}}\circ D_{C_0}{\mathcal{F}}(\bar{I};0),$$
so if this is singular there exists a nontrivial $\tilde{C}_0\in X$
with
\begin{equation}\label{sing}
\tilde{C}_0=-{\mathcal{L}}(
D_{C_0}{\mathcal{F}}(\bar{I};0)(\tilde{C}_0)).
\end{equation}
We will thus assume that (\ref{sing}) holds, and show that it forces
$\tilde{C}_0=0$.

By the definition of the Fr\'echet derivative,
\begin{equation}\label{fre}
D_{C_0}{\mathcal{F}}(\bar{I};0)(\tilde{C}_0)=\frac{d}{d\alpha}{\mathcal{F}}(\bar{I};\alpha
\tilde{C}_0)\Big|_{\alpha=0}.
\end{equation}

By the definition of ${\mathcal{F}}$, we have
\begin{equation}\label{df}{\mathcal{F}}(\bar{I};\alpha \tilde{C}_0)=S_t(\alpha,t),
\end{equation}
where $S(\alpha,t)$ satisfies
\begin{equation}\label{dc}S_t(\alpha,t)+k_1[K-\bar{C}-\alpha\tilde{C}_0(t)]S(\alpha,t)
=\bar{I}+k_{-1}\alpha \tilde{C}_0(t)+k_{-1}\bar{C}
\end{equation}
In particular
$$S(0,t)=\bar{S},$$
where $\bar{S}$ is given in (\ref{stat}). Differentiating (\ref{dc})
with respect to $\alpha$ and setting $\alpha=0$ we get
\begin{eqnarray}\label{dcd}S_{\alpha t}(0,t)+
k_1[K-\bar{C}]S_\alpha(0,t)=[k_1 \bar{S}+k_{-1}]\tilde{C}_0(t)
\end{eqnarray}
From (\ref{fre}) and (\ref{df}) we have
$$D_{C_0}{\mathcal{F}}(I;0)(\tilde{C}_0)=S_{\alpha t}(0,t),$$
which, using (\ref{sing}), implies
$${\mathcal{L}}(S_{\alpha t}(0,\cdot))=-\tilde{C}_0,$$
so by the definition of ${\mathcal{L}}$
\begin{eqnarray}\label{dcd1}
\tilde{C}_0'(t)+k_2 \tilde{C}_0(t)=-S_{\alpha t}(0,t)
\end{eqnarray}
Differentiating (\ref{dcd}) with respect to $t$ we have
$$S_{\alpha tt}(0,t)+
k_1[K-\bar{C}]S_{\alpha t}(0,t)=[k_1
\bar{S}+k_{-1}]\tilde{C}_0'(t)$$ and substituting (\ref{dcd1}) we
have
\begin{equation}\label{xx}S_{\alpha tt}(0,t)-
k_1[K-\bar{C}][\tilde{C}_0'(t)+k_2 \tilde{C}_0(t)]=[k_1
\bar{S}+k_{-1}]\tilde{C}_0'(t).
\end{equation}
Differentiating (\ref{dcd1}) with respect to $t$ we have
$$S_{\alpha tt}(0,t)=-\tilde{C}_0''(t)-k_2 \tilde{C}_0'(t),$$
and together with (\ref{xx}) we get
$$\tilde{C}_0''(t)+[k_2 +
k_1(K-\bar{C}+\bar{S})+k_{-1}]\tilde{C}_0'(t) +k_1k_2[K-\bar{C}]
\tilde{C}_0(t)=0.$$ Multiplying this by $\tilde{C}'_0(t)$ and
integrating over $[0,T]$, taking into account the periodicity, we
obtain
$$[k_2 +
k_1(K-\bar{C}+\bar{S})+k_{-1}]\int_0^T (\tilde{C}_0'(t))^2dt=0,$$
and since, using (\ref{ec1}), the coefficient is positive, we have
$\tilde{C}_0'=0$. Since $\tilde{C}_0\in X$, this implies
$\tilde{C}_0=0,$ as we wanted to prove.
\end{proof}

From lemmas \ref{hi} and \ref{dnz} we obtain
$$deg(id_X-{\mathcal{H}}(1,\cdot),G)\neq 0$$
which implies the existence of a solution $C_0\in G$ of (\ref{ho}),
which, by lemma \ref{nn}, implies theorem \ref{main}.

\medskip

\end{document}